\documentclass[%
 reprint,
superscriptaddress,
 amsmath,amssymb,
prb,
]{revtex4-2}

\usepackage{mhchem}
\usepackage{siunitx}
\usepackage{graphicx}
\usepackage{dcolumn}
\usepackage{bm}
\usepackage{hyperref}
\usepackage{upgreek}

\begin{document}

\preprint{APS/123-QED}

\title{Exploring the interfacial coupling between graphene and the antiferromagnetic insulator MnPSe$_3$}

\author{Xin Yi}
\affiliation
{MOE Key Laboratory of Fundamental Physical Quantities Measurement \& Hubei Key Laboratory of Gravitation and Quantum Physics, PGMF and School of Physics, Huazhong University of Science and Technology, Wuhan 430074, China}

\author{Qiao Chen}
\affiliation
{MOE Key Laboratory of Fundamental Physical Quantities Measurement \& Hubei Key Laboratory of Gravitation and Quantum Physics, PGMF and School of Physics, Huazhong University of Science and Technology, Wuhan 430074, China}

\author{Kexin Wang}
\affiliation
{MOE Key Laboratory of Fundamental Physical Quantities Measurement \& Hubei Key Laboratory of Gravitation and Quantum Physics, PGMF and School of Physics, Huazhong University of Science and Technology, Wuhan 430074, China}

\author{Yuanyang Yu}
\affiliation
{MOE Key Laboratory of Fundamental Physical Quantities Measurement \& Hubei Key Laboratory of Gravitation and Quantum Physics, PGMF and School of Physics, Huazhong University of Science and Technology, Wuhan 430074, China}

\author{Yi Yan}
\affiliation
{MOE Key Laboratory of Fundamental Physical Quantities Measurement \& Hubei Key Laboratory of Gravitation and Quantum Physics, PGMF and School of Physics, Huazhong University of Science and Technology, Wuhan 430074, China}

\author{Xin Jiang}
\affiliation
{MOE Key Laboratory of Fundamental Physical Quantities Measurement \& Hubei Key Laboratory of Gravitation and Quantum Physics, PGMF and School of Physics, Huazhong University of Science and Technology, Wuhan 430074, China}

\author{Chengyu Yan}
\email{chengyu\_yan@hust.edu.cn}
\author{Shun Wang}
 \email{shun@hust.edu.cn}
\affiliation{MOE Key Laboratory of Fundamental Physical Quantities Measurement \& Hubei Key Laboratory of Gravitation and Quantum Physics, PGMF and School of Physics, Huazhong University of Science and Technology, Wuhan 430074, China}
\affiliation{Institute for Quantum Science and Engineering, Huazhong University of Science and Technology, Wuhan 430074, China}


\date{\today}

\begin{abstract}

Interfacial coupling between graphene and other 2D materials can give rise to intriguing physical phenomena. In particular, several theoretical studies predict that the interplay between graphene and an antiferromagnetic insulator could lead to the emergence of quantum anomalous Hall phases. However, such phases have not been observed experimentally yet, and further experimental studies are needed to reveal the interaction between graphene and antiferromagnetic insulators. Here, we report the study in heterostructures composed of graphene and the antiferromagnetic insulator MnPSe$_3$. It is found that the MnPSe$_3$ has little impact on the quantum Hall phases apart from doping graphene via interfacial charge transfer. However, the magnetic order can contribute indirectly via process like Kondo effect, as evidenced by the observed minimum in the temperature-resistance curve between 20-40 K, far below the Néel temperature (70 K).

\end{abstract}

\maketitle


\section{Introduction}

Recent theoretical studies predict that the interaction between graphene and antiferromagnetic materials (AFM) can open a gap at the Dirac point of graphene and induce quantum anomalous Hall effect \cite{qiao2014quantum,zhang2015quantum,takenaka2019magnetoelectric,hogl2020quantum,zhang2018strong}, which renders AFM/graphene heterostructure a promising platform for quantum information and quantum metrology. However, the signature of quantum anomalous Hall effect has not been unveiled in experimental work conducted with a large series of AFM/graphene heterostructure, including  Graphene/CrI$_3$ \cite{tseng2022gate},  Graphene/RuCl$_3$ \cite{mashhadi2019spin,zhou2019evidence}, Graphene/CrOCl \cite{wang2022quantum} and so on \cite{tang2020magnetic,wu2020large,zhang2020mnps}. Rather, the presence of the AFM mainly modulates the width of the quantum Hall plateaus or leads to a noticeable asymmetry in the quantum Hall behavior between the n-doping and p-doping regime \cite{tseng2022gate,wang2022quantum}. Most of the experiments attribute the aforementioned observations to the interfacial charge transfer process rather than the magnetic order of the AFM \cite{tseng2022gate,wang2022quantum}. On the other hand, several works argue that the magnetic order may be crucial due to a minimum in the resistance-temperature curve of the heterostructure occurs at the Néel temperature of the AFM \cite{mashhadi2019spin,zhou2019evidence}. 

Here, we perform the experiments in Graphene/MnPSe$_3$ heterostructure to clarify the role of the magnetic order. Besides, the strong coupling of the spin-valley degree of freedom to antiferromagnetic order in MnPSe$_3$ \cite{li2013coupling} may be an additional knob to realize quantum anomalous Hall effect in such a structure. The dominant role of the interfacial charge transfer is confirmed by both transport measurements and Raman spectroscopy characterizations. However, the contribution of the magnetic order should not be discarded, since we have observed a $R(T)$ minimum between 20-40 K. Our data suggest that the origin of the resistance minimum is not directly associated with the previously reported antiferromagnetic phase transition \cite{mashhadi2019spin,zhou2019evidence}, since it is located far from the Néel temperature. Instead, it might be a direct result of the Kondo effect arising from the local magnetic order in AFM.

\section{Synthesis and Characterization of MnPSe$_3$}

\begin{figure*}[htb]
	\centering
	\includegraphics[scale=0.5]{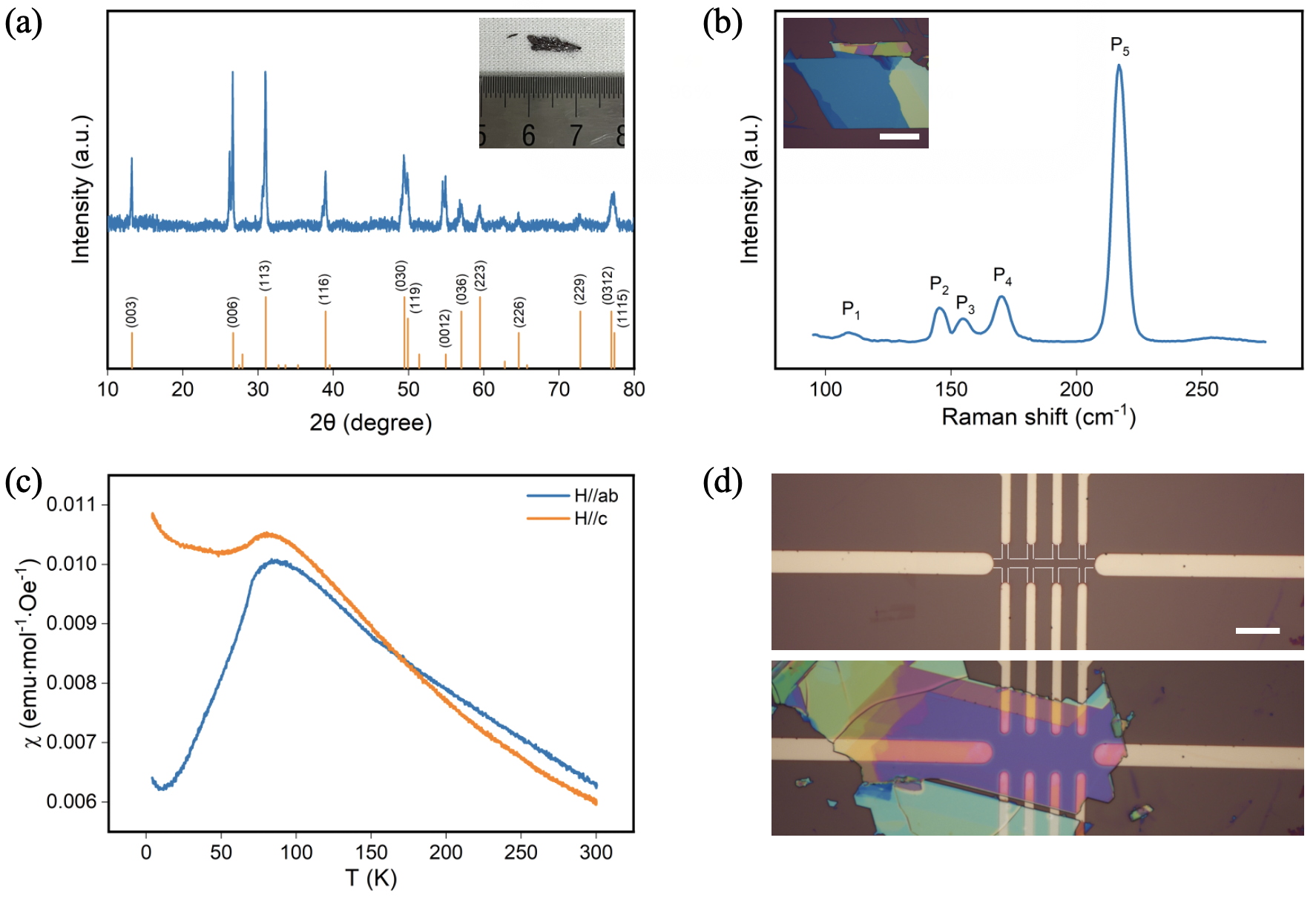}
	\caption{\label{fig1} (a) The XRD characterization results of MnPSe$_3$ powder (blue curve). The red vertical line marks the standard MnPSe$_3$ pattern (PDF \#33-0902). The inset shows the grown MnPSe$_3$ crystal. (b) Optical image (inset) and Raman spectra of the MnPSe$_3$ flake. (c) The temperature dependence of magnetic susceptibility of MnPSe$_3$ single crystal under an external magnetic field of 1000 Oe for H//ab and H//c, respectively. (d) Optical image of the graphene device before (top) and after (bottom) being covered by MnPSe$_3$. Scale bar: (b) 20 $\upmu$m, (d) 10 $\upmu$m.
	}
\end{figure*}

MnPSe$_3$ crystals were synthesized via the chemical vapor transport (CVT) method. Raw materials, including manganese, phosphorus, and selenium, were mixed in a stoichiometric ratio of 1:1:3 and sealed in an evacuated quartz tube with iodine powder as a transport agent. The tube was then placed in a double-zone furnace, with the hot end at 800 $^{\circ}$C and cold end at 700 $^{\circ}$C, creating a temperature gradient along the tube. After a growth period of 14 days, the quartz tube was cooled to room temperature naturally, resulting in the formation of MnPSe$_3$ crystals. The inset of Figure~\ref{fig1}a displays an optical image of MnPSe$_3$ crystals grown by the CVT method. The close match between then X-ray diffraction (XRD) and standard MnPSe$_3$ pattern indicates the purity of the sample. A Raman spectroscopy characterization at 300 K with a laser excitation wavelength of 488 nm was conducted after exfoliating the MnPSe$_3$ flake onto a SiO$_2$/Si substrate (Figure~\ref{fig1}b). The characteristic peaks of MnPSe$_3$, labeled as P$_1$-P$_5$, were observed at 110 cm$^{-1}$, 145 cm$^{-1}$, 155 cm$^{-1}$, 170 cm$^{-1}$, and 217 cm$^{-1}$, respectively, consistent with other reports \cite{liu2020exploring,liu2021synthesis}. The magnetic susceptibility of MnPSe$_3$ was characterized using a Vibrating Sample Magnetometer (Quantum Design) in the presence of a magnetic field of 1000 Oe. The MnPSe$_3$ crystal responded differently to the magnetic field direction as shown in Figure~\ref{fig1}c, indicating the magnetic anisotropy \cite{liu2020exploring,jeevanandam1999magnetism}. The Néel temperature of 70 K was determined from the maximum slope of the susceptibility curve (H//ab).

\section{Transport measurements}

\begin{figure*}[htb]
	\centering
	\includegraphics[scale=0.5]{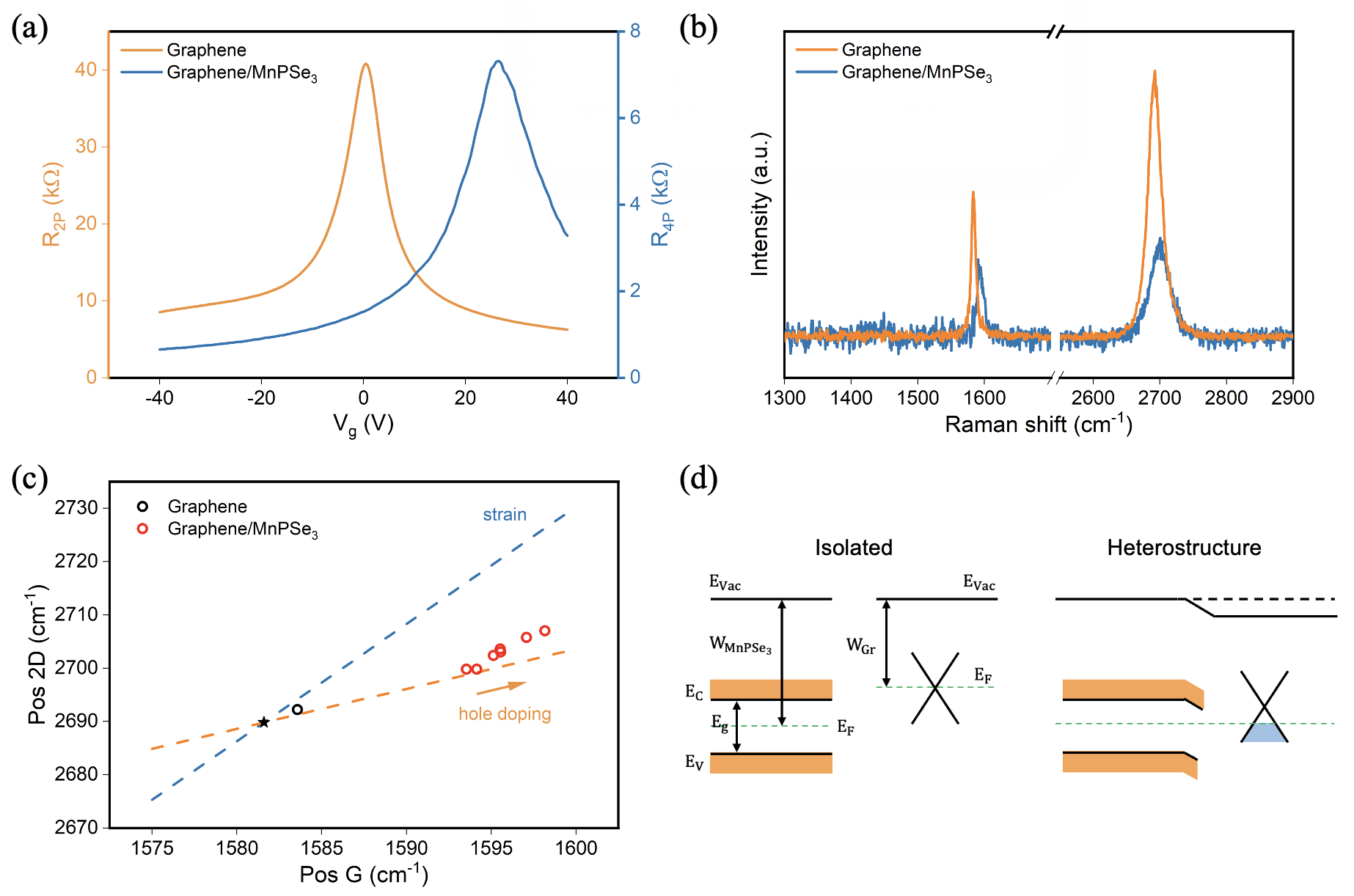}
	\caption{\label{fig2} (a) Longitudinal resistance as a function of gate voltage in graphene and graphene/MnPSe$_3$ devices at 300 K. The vertical axis $R_{2P}$ and $R_{4P}$ represent the two-terminal and four-terminal resistances, respectively. (b) Raman spectra of graphene before and after transfer of MnPSe$_3$. (c) Positions of G and 2D peaks of graphene at different locations of the Graphene/MnPSe$_3$ heterojunction (red circles).  The black circle indicates the measurement result of graphene before transfer of MnPSe$_3$ in (b). Blue and orange dashed lines represent the frequency shift of the G and 2D peaks of graphene under stress and doping, respectively. The black star, situated at the intersection of two dashed lines, denotes the position of G and 2D peaks of graphene in the absence of strain or doping. (d) Schematic of energy levels for MnPSe$_3$ and graphene before and after contact. Here, $W_{MnPSe_3}=6.23\ \rm eV$ and $E_g=2.54\ \rm eV$ are the work function and bandgap of MnPSe$_3$, and $W_{Gr}=4.29\ \rm eV$ is the work function of graphene.
	}
\end{figure*}

After characterizing the MnPSe$_3$, we employed a dry transfer technique \cite{yi2022quantum} to transfer the MnPSe$_3$ flake onto the surface of the graphene device (Figure~\ref{fig1}d). To ensure the integrity of the transferred MnPSe$_3$, MnPSe$_3$  flakes with thicknesses ranging from 10 to 50 nm were utilized. Four devices marked as D1-D4 are measured, they show similar behavior. The main results are rendered from device D1 unless specified otherwise. A complete data set for D2 can be found in the supplemental material. The presence of the MnPSe$_3$ flakes modulated the doping level of the graphene noticeably, shifting the Dirac points with respect to the back-gate voltage $V_g$ from $V_g=0\ \rm V$ (orange curve in Figure~\ref{fig2}a) to $V_g=26\ \rm V$ (blue curve). Based on the thickness of the SiO$_2$ substrate (285 nm), the doping level is $1.97\times10^{12}$ cm$^{-2}$. To further characterize the doping effect, we performed Raman spectroscopy characterizations on graphene before and after transferring the MnPSe$_3$ flake. An exemplified result is shown in Figure~\ref{fig2}b, it is found that both the G and 2D peaks showed red shifts after the graphene got covered by the MnPSe$_3$ flake. The shift of the G and 2D peaks (Pos G, Pos 2D) of graphene can be induced by either strain or interfacial charge transfer \cite{yang2019all,lee2012optical,zhang2019study}. Strain affects the bond lengths, while charge transfer caused doping influences the electron-phonon coupling. Both of these factors contribute to variations in the phonon frequencies of graphene, which can be distinguished by the ratio $\Delta$Pos 2D/$\Delta$Pos G, where $\Delta$Pos 2D is shift of the 2D peak and $\Delta$Pos G is that of the G peak. The ratio $\Delta$Pos 2D/$\Delta$Pos G is 2.2 for the strain scenario and 0.75 for the case of interfacial charge transfer \cite{yang2019all,lee2012optical,zhang2019study}. A series of Raman characterizations were carried out at different locations of the graphene/MnPSe$_3$ heterojunction to determine $\Delta$Pos 2D/$\Delta$Pos G and hence distinguish the contributions of the two effects. Combining all the results together, as shown in Figure~\ref{fig2}c, it is obvious $\Delta$Pos 2D/$\Delta$Pos G agrees well the prediction by interfacial charge transfer effect. Besides, the scatter in the results acquired at different sample locations suggest the charge transfer and thereby the doping level may not be uniform throughout the sample. Also, the results may vary from sample to sample (Figure~\ref{fig2}a and Figure S1b) due to the inhomogeneous contact between graphene and MnPSe$_3$.

\begin{figure*}[htb]
	\centering
	\includegraphics[scale=0.5]{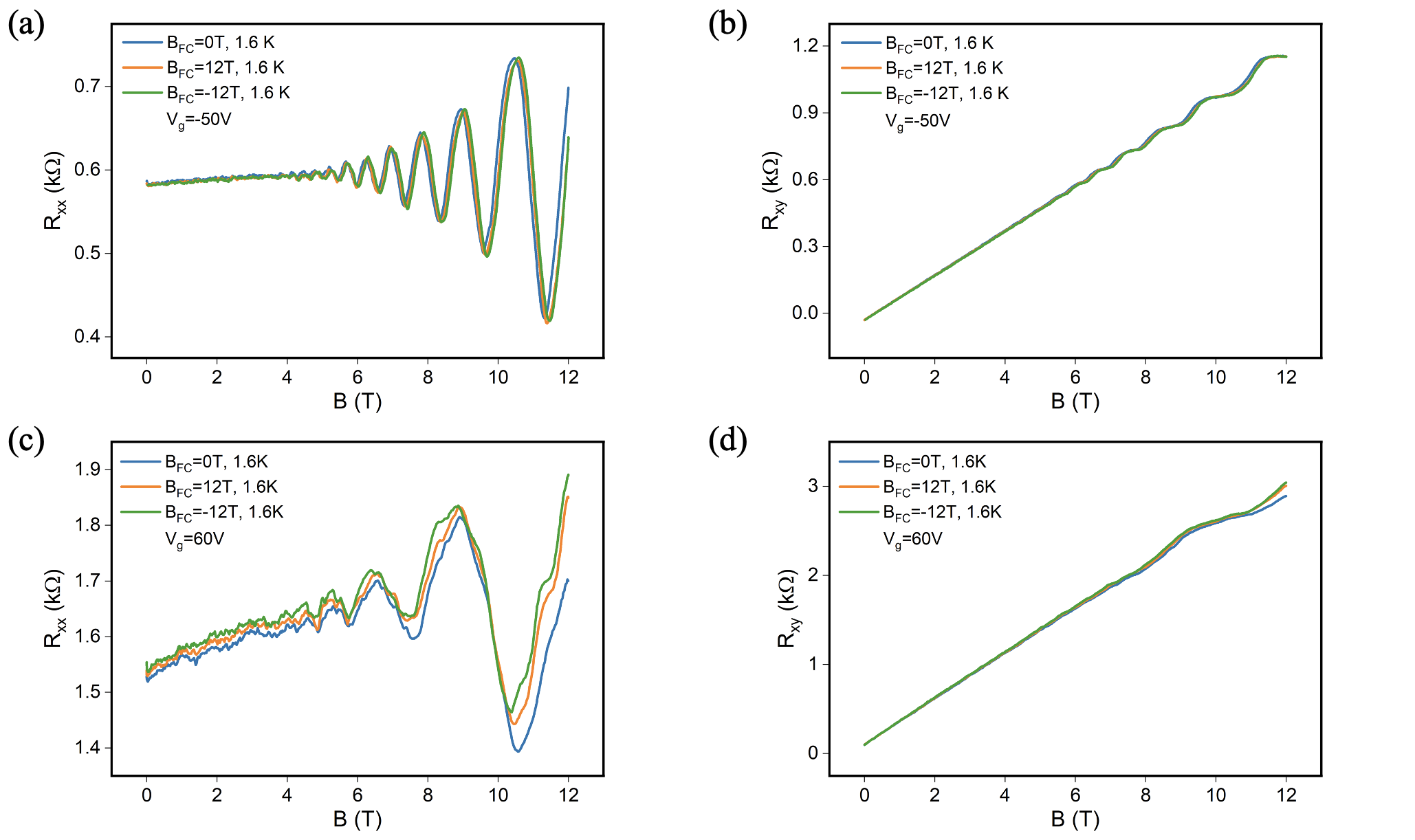}
	\caption{\label{fig3} Quantum oscillations and quantum Hall effect under different gate voltages and cooling fields. (a) and (b) depict the quantum oscillations and quantum Hall effect observed at different cooling field ($B_{FC}$) when $V_g=-50\ \rm V$ (p-doping), showing no drift. (c) and (d) show that the quantum oscillations exhibit small drift of about 100-300 mT under different cooling field when $V_g=60\ \rm V$ (n-doping), with no significant effect on the quantum Hall effect.
	}
\end{figure*}

\begin{figure*}[htb]
	\centering
	\includegraphics[scale=0.5]{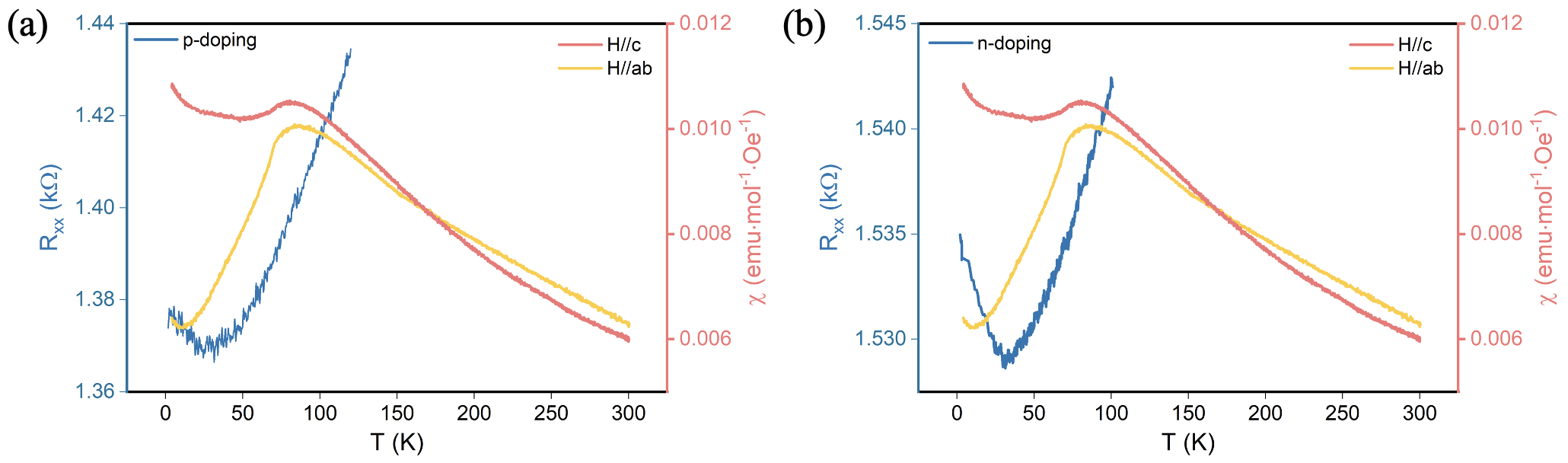}
	\caption{\label{fig4} Temperature dependence of longitudinal resistance measured at $V_g=0\ \rm V$ (p-doping) (a) and $V_g=60\ \rm V$ (n-doping) (b), respectively. The temperature dependence of magnetic susceptibility curve corresponding to the right y-axis is identical to Figure~\ref{fig1}c.
	}
\end{figure*}

\begin{figure*}[htb]
	\centering
	\includegraphics[scale=0.5]{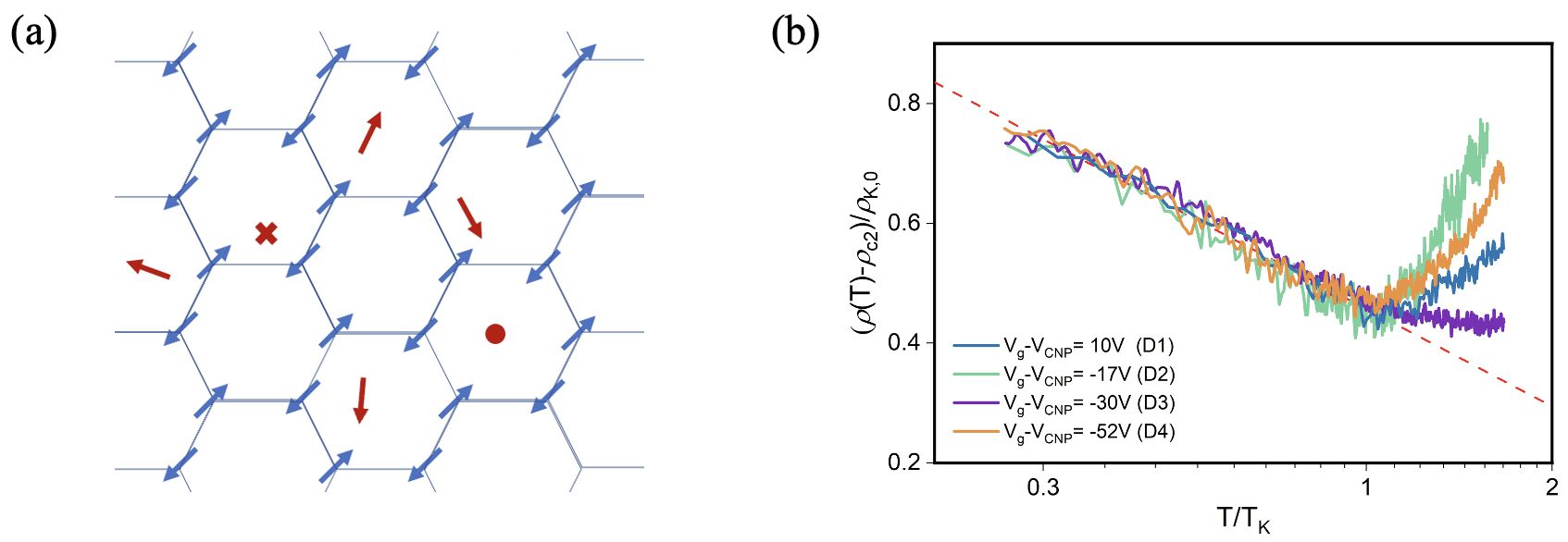}
	\caption{\label{fig5} (a) Magnetic structure of MnPSe$_3$. The red arrow, circle, and cross represent the magnetic moments of impurities pointing in different directions. (b) Fitting and normalization of the $R-T$ curve using the Kondo theory. The four different colored curves represent the experimental data obtained in four different devices, covering both n-doping and p-doping cases. The red dashed line shows the prediction of Kondo effect (equation~(\ref{eq:two})).
	}
\end{figure*}

To better understand the charge transfer process, we analyzed the band alignment at the interface of Graphene/MnPSe$_3$ heterostructure, taking into account the work functions of both materials. Figure~\ref{fig2}d depicts a schematic representation of the energy levels for MnPSe$_3$ and graphene before and after contact. The work function of MnPSe$_3$ ($W_{MnPSe_3}$), the work function of graphene ($W_{Gr}$), and the band gap of MnPSe$_3$ ($E_g$) were 6.23 eV, 4.29 eV and 2.54 eV, respectively \cite{zhang2018strong,sharma2020strain}. When these two materials form a heterostructure, the conduction band and valance band edge of MnPSe$_3$ will bend downward, the valance band electrons in graphene will populate the empty conduction band of MnPSe$_3$. The graphene will be p-doped due to the charge transfer process.

Apart from the doping effect, the role of the magnetic order of MnPSe$_3$ is a much more crucial topic. It is proposed to be the key ingredient to drive a quantum anomalous Hall state. To test the role of the magnetic order, we studied the quantum Hall effect with magnetic field cooling protocol.

First of all, the sample is cooled in the absence of a magnetic field and then characterized at 1.6 K ($B_{FC}=0\ \rm T$ traces in Figure~\ref{fig3}). The longitudinal and Hall resistances as a function of the magnetic field at $V_g=-50\ \rm V$ and $V_g=60\ \rm V$, respectively, revealing pronounced quantum oscillations and Hall plateaus. Then, the sample were cooled in the presence of a magnetic field and followed by demagnetization at low temperature. It is expected that the field cooling can modulate the interfacial coupling by manipulating the Néel vector \cite{ambrose1998dependence,koon1997calculations,hoffmann2004symmetry}, thereby introducing a spin splitting into graphene and shifting in the Landau levels \cite{wu2020large}. However, this shift is absence in our study in both n-doped and p-doped regime, the main features remain the same as those measured without field cooling. These results indicate that MnPSe$_3$ did not modify the band structure or spin pattern of graphene.

However, it is still interesting to study if the magnetic order of MnPSe$_3$  can affect the transport indirectly via processes such as weak localization or Kondo effect. The magnetoresistance in the small field regime suggests graphene/MnPSe$_3$ behaves similarly to the bare graphene, and thus excludes the possibility that the presence of MnPSe$_3$ may affect weak localization (Figure S4). The temperature dependence of resistance, on the other hand, reveals interesting results. The temperature-resistance curve ($R-T$ curve) of the heterostructure shows a non-monotonic behavior, i.e., the resistance first drops with increasing temperature up to 20-40 K (depends on the doping level) and then rises all the way to high temperature, as shown in Figure~\ref{fig4}. Such phenomenon was not observed in bare graphene devices (Figure S3). The observed resistance minimum, located far below the Néel temperature (70 K), is a significant deviation from previous reports on other graphene/AFM heterostructures, where the minimum is closer to the Néel temperature \cite{mashhadi2019spin,zhou2019evidence}. The most probable explanation for the observed phenomenon is the presence of magnetic impurities (Figure~\ref{fig5}a) or imperfection of the crystal lattice, as evidenced by the slight increase in magnetic susceptibility at temperatures below the Néel temperature \cite{isobe2002novel,petrova2005spin}. Furthermore, due to the significant magnetic anisotropy of MnPSe$_3$ \cite{jeevanandam1999magnetism,liu2020exploring}, the increase in magnetic susceptibility is more pronounced along the c-axis. The magnetic impurities could potentially induce the Kondo effect in graphene, leading to a minimum in the $R-T$ curve \cite{chen2011tunable}. To verify this conjecture, the experimental data were fitted using theories of the Kondo effect. The influence of the Kondo effect on the resistivity in the low-temperature and intermediate-temperature ranges are expected to behave, respectively, as \cite{nozieres1974fermi,costi1994transport,chen2011tunable}

\begin{eqnarray}
	\rho(T)=\rho_{\mathrm{c} 1}+\rho_{K, 0}\left(1-\left(\frac{T}{T_K}\right)^2\right) \label{eq:one}
	\\
	\rho(T)=\rho_{\mathrm{c} 2}+\frac{\rho_{K, 0}}{2}\left(1-0.47 \ln \left(\frac{1.2 T}{T_K}\right)\right) \label{eq:two}
\end{eqnarray}

where $\rho_{c1}$ and $\rho_{c2}$ are the non-temperature dependent parts of the resistivity, and they can be extracted from the $R-T$ curve in the metallic regime (Figure S5), $\rho_{K,0}$ is the Kondo resistivity at zero temperature, and $T_K$ is the Kondo temperature. Figure~\ref{fig5}b highlights the resistance upturn behavior, arising from the Kondo effect, can be observed in 4 devices with different MnPSe$_3$ thickness. The Kondo temperature spans in the range of 30-40$\pm$10 K. The relatively large uncertainty in $T_K$ is primarily due to the lacking of the low temperature data (the base temperature of our cryostat is 1.6 K). We expect a joint fitting with equation~\ref{eq:one} and~\ref{eq:two} can squeeze the uncertainty in $T_K$ should the cryostat can go to lower temperature. $T_K$ can cover a large temperature range in graphene, 10-90 K \cite{chen2011tunable}, therefore it should be extremely cautious to attribute the resistance minimum or other resistance anomalies to AFM phase transition even when the minimum occurs near the Néel temperature.

\section{Conclusion}

In summary, we have investigated the electronic transport properties of graphene devices in the presence of a nearby MnPSe$_3$ flake. Our findings, which combining transport measurements with Raman spectroscopy, provide compelling evidence for charge transfer between graphene and MnPSe$_3$. Although our results indicates the magnetic order of MnPSe$_3$ do not have direct contribution to the band structure or spin pattern in graphene, it can have indirect impact on the transport properties via process like Kondo effect. The latter is demonstrated by the presence of a resistance minimum between 20-40 K, as shown in the temperature-resistance curve, far below the Néel temperature (70 K). This highlights the feasibility of realizing the spintronic device in graphene/AFM heterostructure even in the absence of quantum anomalous Hall effect.

\begin{acknowledgments}
We thank professor Zhaoming Tian for magnetic susceptibility discussion. This work is supported by the National Natural Science Foundation of China with Grants Numbers 12074134, 12204184 and U20A2077.
\end{acknowledgments}

\nocite{*}

\providecommand{\newblock}{}

\end{document}